%% file: MainDoc.tex
\documentclass[a4paper, amsfonts, amssymb, amsmath, twocolumn, showkeys, nofootinbib,superscriptaddress]{revtex4-1}

\input{preamble}
\usepackage[english]{babel}
\usepackage[utf8]{inputenc}
\usepackage[colorinlistoftodos, color=green!40, prependcaption]{todonotes}
\input{preamble}
\usepackage[pdftex, pdftitle={Article}, pdfauthor={Author}]{hyperref} 

\begin{document}

\title{Silicon oxynitride platform for linear and nonlinear photonics at NIR wavelengths}

\author{Gioele Piccoli}
\email{gpiccoli@fbk.eu}
\affiliation{Sensors and Devices, Fondazione Bruno Kessler, Trento, Italy}
\affiliation{Department of Physics, University of Trento, Trento, Italy}

\author{Matteo Sanna}
\affiliation{Department of Physics, University of Trento, Trento, Italy}

\author{Massimo Borghi}
\affiliation{Department of Physics, University of Trento, Trento, Italy}
\affiliation{Present address: Department of Physics, University of Pavia, Pavia, Italy}

\author{Lorenzo Pavesi}
\affiliation{Department of Physics, University of Trento, Trento, Italy}

\author{Mher Ghulinyan}
\affiliation{Sensors and Devices, Fondazione Bruno Kessler, Trento, Italy}


\begin{abstract}
The development of versatile and novel material platforms for integrated photonics is of prime importance in the perspective of future applications of photonic integrated circuits for quantum information and sensing. Here we present a low-loss material platform based on high-refractive index silicon oxynitride (\ce{SiON}), which offers significant characteristics for linear and non-linear optics applications in a wide range of red/near-infrared wavelengths. The demonstrated propagation loss $<1.5$~dB/cm for visible wavelengths enables the realization of long and intricate circuitry for photon manipulations, as well as the realization of high quality factor resonators. In addition, the proposed \ce{SiON} shows a high nonlinear coefficient of $10^{-19}$~m$^2$/W, improving the strength of nonlinear effects exploitable for on-chip photon generation schemes.
\end{abstract}

\maketitle


\section{Introduction}
\label{sec-intro}
The development of novel platforms for integrated photonics, characterized by a strong versatility in terms of applications, is of prime importance for the realization of emerging photonics applications in the fields of sensing, metrology, quantum communications and quantum computing~\cite{capmany2020programmable,osgood2021principles}. 
Several technologies for photonic integrated circuits (PICs) have been proposed and developed in the last decades. 
Silicon photonics emerged first, exploiting the advanced CMOS manufacturing techniques optimized for microelectronics. Nowadays, \ce{Si} photonic devices, based on the Silicon-on-Insulator (SOI) platform, are already spreading in the commercial world, mainly for the telecommunication networks (see Ref.~\cite{siew2021review} and references therein). State of the art commercial SOI devices, designed for the telecom C-band, achieve propagation losses of less than $1$~dB/cm, offering highly efficient passive components distributed within standard product development kit~\cite{bogaerts2018silicon,fahrenkopf2019aim,aalto2019open}. 

Active components are also implemented within the SOI platform, with phase shifters based on \textit{p-n} junctions realized directly on the \ce{Si} waveguides, while for photodetectors and light sources hetero-integration of germanium or III-V compound semiconductor materials is employed. A limiting constrain of \ce{Si} PICs is that their linear operation is acceptable at wavelengths longer than $1.1~\mu$m (energy bandgap $E_g\sim1.1$~eV) where the core material's absorption is insignificant, while for nonlinear optical applications excited-carrier (EC) and two-photon absorption (TPA) can still be significant at photon energies $\sim E_g/2$. 

Silicon Nitride (\ce{SiN}) has been introduced as an alternative dielectric platform for integrated photonics~\cite{blumenthal2018silicon}. It is transparent at wavelengths from $400$~nm to $2.0 \mu$m, while very recent developments report on operation in the ultra-violet region~\cite{morin2021cmos}.
Commercially available platforms show propagation losses $<0.1$~dB/cm over the whole operational range, with passive components that show similar performance to their SOI counterparts~\cite{moss2013new,roeloffzen2018low,liu2021high}. Due to its dielectric nature, active components in \ce{SiN} technology can relay to only relatively slower thermo-optical modulation of the refractive index. Nevertheless, due to the large optical bandgap of $E_g >4$~eV, \ce{SiN} does not suffer EC or TPA losses, while nonlinear optical generation can still be significant owing to appreciable third-order nonlinearities. 
Stoichiometric silicon nitride (\ce{Si3N4}) offers superior optical quality, however, the film thicknesses should be kept to $<200$~nm to avoid film cracking due to the large tensile stress. This limitation can be overcome using  sophisticated techniques, such as photonic Damascene~\cite{pfeiffer2016photonic} and multilayered TriPleX~\cite{heideman2005low,morichetti2007box,roeloffzen2018low} processes.

An alternative approach to reduce film stress consists in introducing oxygen into the \ce{SiN} material, by depositing directly a silicon oxynitride (\ce{SiON}) film~\cite{worhoff1999plasma,germann2000silicon,samusenko2016sion,chen2017onchip}. The refractive index of \ce{SiON} can be continuously tuned between 1.45 (\ce{SiO2}) and 2.00 (\ce{Si3N4}) by controlling the relative content of \ce{O} and \ce{N} in the film. The material loss can be lower than that of \ce{SiN}, meantime maintaining a very low stress for film thicknesses of up to few micrometers. As a drawback, the reduction of the material refractive index causes an increase in the device footprint, as well as a weakening of the thermo-optical and nonlinear optical characteristics of the guiding components~\cite{trenti2018thermopt}.

\begin{figure*}[htb!]
\centering\includegraphics[width=0.66\textwidth]{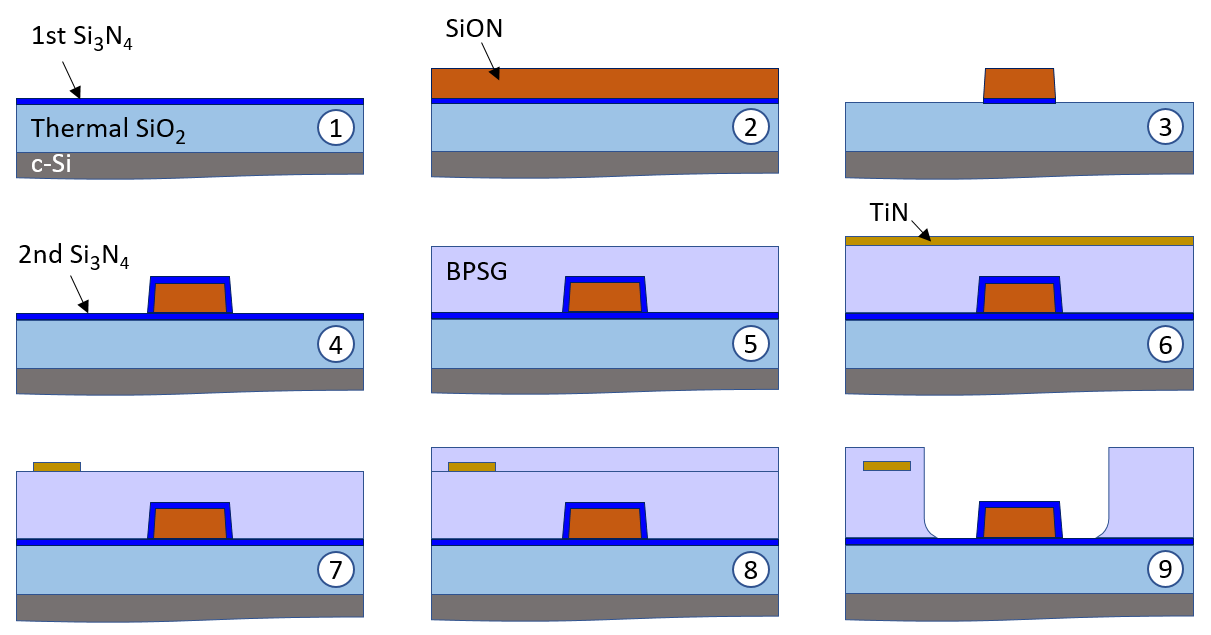}
\caption{Sketch of the fabrication process of the high-index \ce{SiON} photonic platform. (1)~The \ce{Si} wafer with the bottom \ce{SiO2} cladding and the \ce{Si3N4} film, (2)~deposition of the \ce{SiON} core material, (3)~lithography and etching of waveguides, (4)~deposition of the top \ce{Si3N4} film, (5)~deposition of the top \ce{SiO2} cladding, (6)~sputtering of the \ce{TiN} film, (7)~lithography and etching of the \ce{TiN} resistors, (8)~deposition of an oxidation protective film and (9)~optional local opening of the waveguides top cladding.}
\label{Fig:F1_procflow}
\end{figure*}

In this work, we present a novel, low-loss, photonic platform based on \ce{SiON} channel waveguides, capable to manage a wide range of VIS/NIR wavelengths for both linear and nonlinear optics applications. The core material consists in a relatively high refractive index \ce{SiON} (1.66 at 850nm wavelength), which enables small footprint PIC designs with propagation losses $\sim1.5$~dB/cm, improvable by at least a factor of two. The material shows an optical bandgap of 3.8~eV with a relatively strong optical nonlinearity of $1.3\pm0.6 \times 10^{-19}$~m$^2$/W close to the TPA absorption edge. This, combined with the possibility to remove locally the cladding without damaging the waveguide, offers large versatility to engineer the waveguide dispersion for applications such as nonlinear Four-Wave Mixing (FWM) in ring resonator devices. Our \ce{SiON} platform has the potential to be further developed for the monolithic integration of all necessary functionalities -- photon sources, light manipulation circuits and photon detection (recently proven in Ref.~\cite{bernard2021top}) -- on a single chip, operating at room temperature, for classical and quantum applications.

The paper is structured as follows: Sec.~\ref{sec-platform} presents the technological approach, Sec.~\ref{sec:linprop} and~\ref{sec:nlinprop} describe the linear and nonlinear properties of the \ce{SiON} platform, respectively. In Sec.~\ref{sec:indexeng} we describe the dispersion engineering of ring resonators for non-linear FWM and the fabrication of these devices. Finally, in Sec.~\ref{sec:concl} we summarize our results and draw conclusions.


\section{Photonic platform}
\label{sec-platform}
The photonic platform we introduce is based on the use of high-index \ce{SiON} for the core material and \ce{SiO2} claddings, resulting in a relatively large core/cladding index contrast of $\sim 15\%$. In addition, the channel waveguide is encapsulated between two thin films of \ce{Si3N4}. These last act as an etch-stop barrier during the wet chemical etching, which is used to open windows in the waveguide's top \ce{SiO2} cladding in specific locations on the chip, without the risk to underetch the bottom \ce{SiO2} cladding. The removal of top \ce{SiO2} is an optional process, which can be used for different purposes, for example, a chemical functionalization of the waveguide surface for sensing applications~\cite{densmore2009silicon,washburn2009label,mukundan2009waveguide,heideman2012triplex,samusenko2016sion} or, as it will be discussed in Section~\ref{sec:indexeng}, in cases when the refractive index dispersion of waveguides should be engineered for non-linear optics applications~\cite{turner2006tailored,mas2010tailoring,zhao2015visible,zhang2018temperature,guo2018experimentally,zhao2020visible}.

The fabrication process of the proposed platform is schematically described in Fig.~\ref{Fig:F1_procflow}.
Starting from a 6~inch silicon wafer, first a $1.6~\mu$m silicon oxide has been grown by wet thermal oxidation at $975~^{\circ}$C to form the bottom cladding. On top of it, a $50$~nm thick \ce{Si3N4} is deposited in a Low-pressure chemical vapor deposition (LPCVD) furnace at $770~^{\circ}$C, followed by a $500$~nm \ce{SiON} film deposition in a plasma-enhanced chemical vapor deposition (PECVD) chamber using \ce{SiH4}, \ce{N2O} and \ce{NH3} gas precursors.
Next, the photonic devices were defined by photoresist patterning using an i-line stepper lithography, and the pattern transferred  to the \ce{SiON}/\ce{Si3N4} layers using reactive ion etching (RIE).
Then, the \ce{SiON} waveguides were thermally treated at $1050~^{\circ}$C for 90 min in a \ce{N2} atmosphere to allow the release of residual \ce{H2} and the improvement of the optical properties of the \ce{SiON} film.
Next, a second deposition of $50$~nm \ce{Si3N4} was performed, followed by a deposition of LPCVD borophosphosilicate glass (\ce{BPSG}) and PECVD \ce{SiO2} films at $640~^{\circ}$C and $300~^{\circ}$C, respectively, to form the top cladding of a total of $1.6~\mu$m. 

\begin{figure*}[htb!]
\centering\includegraphics[width=0.66\textwidth]{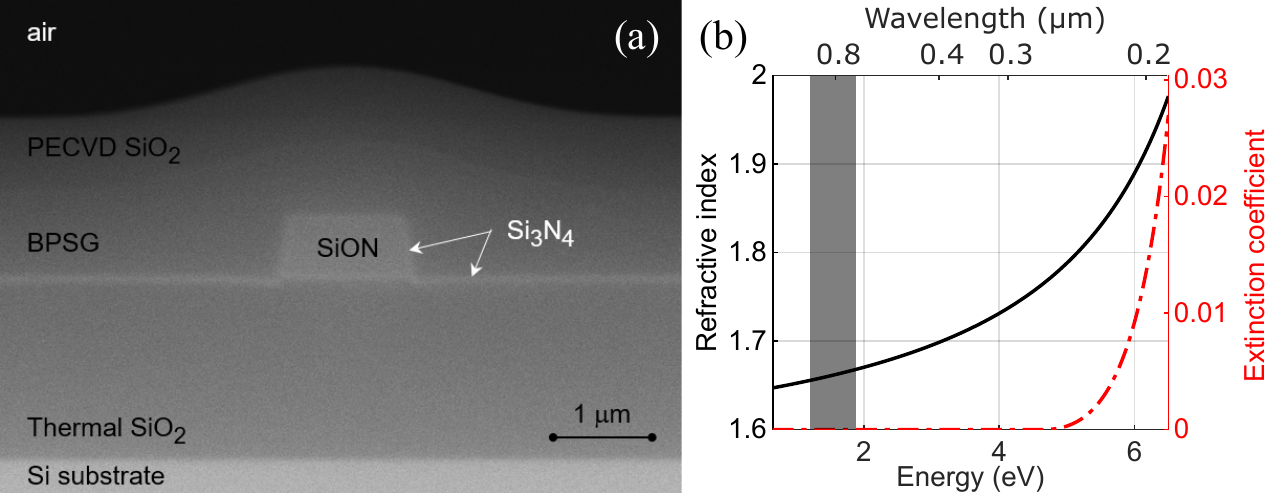}
\caption{(a)~Cross-sectional SEM micrograph of an $800~nm\times 300$~nm \ce{SiON} waveguide. The different materials of the multilayer composing the platform are labelled on the figure. (b)~Dispersions for the real refractive index and the extinction coefficient of the \ce{SiON} core material. The wavelength range of study for this work is highlighted in grey.}
\label{Fig:F12_crossindex}
\end{figure*}

In a next step, a multi-stack of 150~nm \ce{TiN} and 1200~nm of \ce{Al} was sputtered, patterned, and etched with RIE to allow for the realization of metal lines, contact pads and micro-heaters for the thermo-optical tuning of photonic components. The top \ce{Al} film was selectively removed on top of the micro-heaters via wet chemical etching in order to realize efficient \ce{TiN} micro-resistances with a sheet resistance of $5~\Omega$/sq. The wafer was covered with 500~nm PECVD \ce{SiO2} protective film, which was then removed from pad positions to allow for external electrical contact.

Finally, the chip boundaries and the waveguide facets were defined by RIE of the dielectric multilayer, with an additional  $140~\mu$m deep etch into the \ce{Si} substrate through a Bosch process in order to ease the butt-coupling between optical fibers and waveguides.
Figure~\ref{Fig:F12_crossindex}a shows the SEM cross-sectional micrograph of the core of an $800~nm\times 300$~nm \ce{SiON} waveguide. All the different films, including the surrounding $50$~nm-thick \ce{Si3N4} film, the bottom and top claddings as well as the \ce{Si} substrate are clearly visible. 


\section{Linear properties}
\label{sec:linprop}
\subsection{Material dispersion}
The optical properties of the \ce{SiON} films were characterized by variable-angle spectroscopic ellipsometry (VASE). The refractive index was modeled with the \textit{New Amorphous} model based on the Forouhi-Bloomer dispersion equations~\cite{forouhi1988optical}
\begin{equation}
    n(\omega) = n_\infty + \frac{B\cdot (E-E_j)+C}{(E-E_j)^2+\Gamma^2},
\end{equation}

\noindent where $n_\infty$ is the refractive index when $\omega\rightarrow\infty$, $B_j$ and $C_j$ contain the material characteristics
\begin{align}
    B &= \frac{f_j}{\Gamma_j} \cdot ( \Gamma_j^2 - (E_j - E_g)^2 ), \\
    C &= 2 f_j \Gamma_j (E_j - E_g).
\end{align}

\noindent This model allows to estimate the optical band gap $E_g$ of the platform's core material. The parameters $f_j$, $\Gamma_j$ and $E_j$ describe, respectively, the amplitude, the spectral width and the spectral position of the absorption peak in energy units. The obtained dispersions for the real refractive index and the extinction coefficient are plotted in Fig.~\ref{Fig:F12_crossindex}b.

Of particular interest for this work, the refractive index of  \ce{SiON} has been engineered  to be higher than that of typical low-index \ce{SiON} used in previous works~\cite{worhoff1999plasma,bona2003sion,chen2019normal}. This choice allows to design optical components with smaller footprint and larger optical mode confinement. Despite the relatively higher refractive index (1.66 at 850~nm), the estimated band-gap of $E_g \approx 3.8$~eV allows the photonic platform to be used without material absorption loss in the whole near-infrared and visible regions. 
 
\begin{figure*}[htb!]
\centering\includegraphics[width=0.66\textwidth]{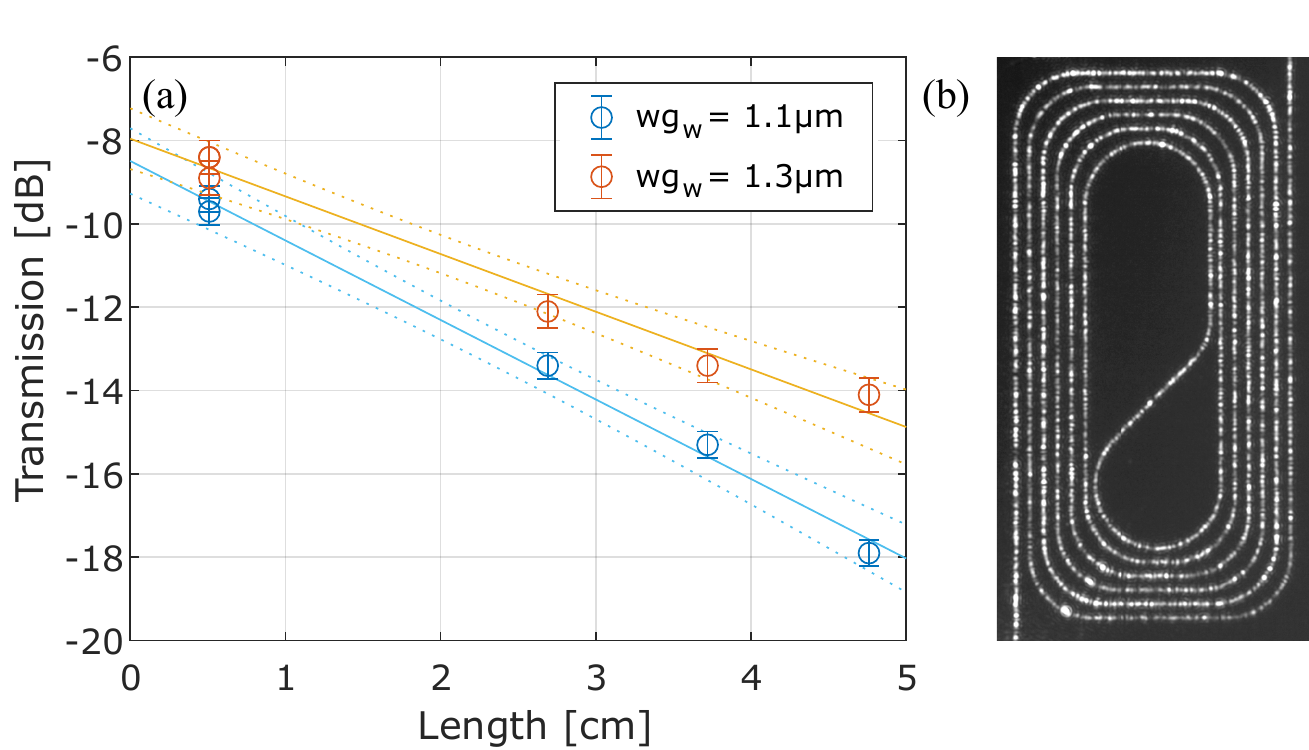}
\caption{(a) Measurement of the propagation losses $\alpha$ through the cutback method at $780$~nm. Two different waveguide widths (wg$_w$), $1.1~\mu$m (blue) and  $1.3~\mu$m (orange) show $\alpha=1.8\pm0.1$~dB/cm and $1.5\pm0.1$~dB/cm, respectively. Errorbars and dotted lines represent the standard deviation of the experimental data and the fit, respectively. (b) Optical image of a $2.7$~cm long spiral waveguide under laser excitation.}
\label{Fig:F3_loss}
\end{figure*} 
\subsection{Propagation loss}
Another important property to be optimized for a photonic platform is the propagation loss of waveguides, that can be attributed to several sources such as: the material absorption, the radiative loss at bends, the loss toward the silicon substrate and the scattering due to surface roughness. The overall value of propagation loss, per unit length, can be characterized emulating the cutback method, by measuring the input/output power ratio of spiral-like waveguides of different lengths, as shown in Fig.~\ref{Fig:F3_loss}.

This characterization was realized for two different widths of waveguides, namely $1.1~\mu$m and $1.3~\mu$m, resulting in average losses of, respectively, $1.8\pm0.2$~dB/cm and $1.5\pm0.2$~dB/cm in the studied wavelength region 740~nm to 840~nm, with a coupling loss of $3.8\pm0.3 $~dB per facet. Note that, at the moment of writing this manuscript, an improvement of our etching technique has lead to propagation losses of the same \ce{SiON} waveguides as low as 0.8 dB/cm.


\section{Non-linear properties}
\label{sec:nlinprop}
The third-order optical nonlinearities of the \ce{SiON} photonic platform have been studied by exploiting the phenomenon of Self-Phase-Modulation (SPM)~\cite{stolen1978self,tzoar1981self}. 
An intense laser pulse, which propagates in a nonlinear medium,  induces a local variation of the refractive index due to strong light-matter interactions. This variation causes a phase shift between the spectral components of the pulse resulting in a modulation of pulse spectrum.
Consequently, by measuring the spectral broadening of an ultra-short pulse with known power, one can retrieve the non-linear index of refraction $n_2$ of the material. In particular, following the split-step method described in Ref.~\cite{agrawal2000nonlinear}, one can simulate the expected SPM effect for a given set of parameters, including: the material's $n_2$ coefficient, the waveguide geometry and the initial characteristics of the pulse. Then, by comparing the simulated results with the measured SPM spectra, the nonlinear coefficient of the material can be estimated.

\begin{figure*}[htb!]
\centering\includegraphics[width=0.66\textwidth]{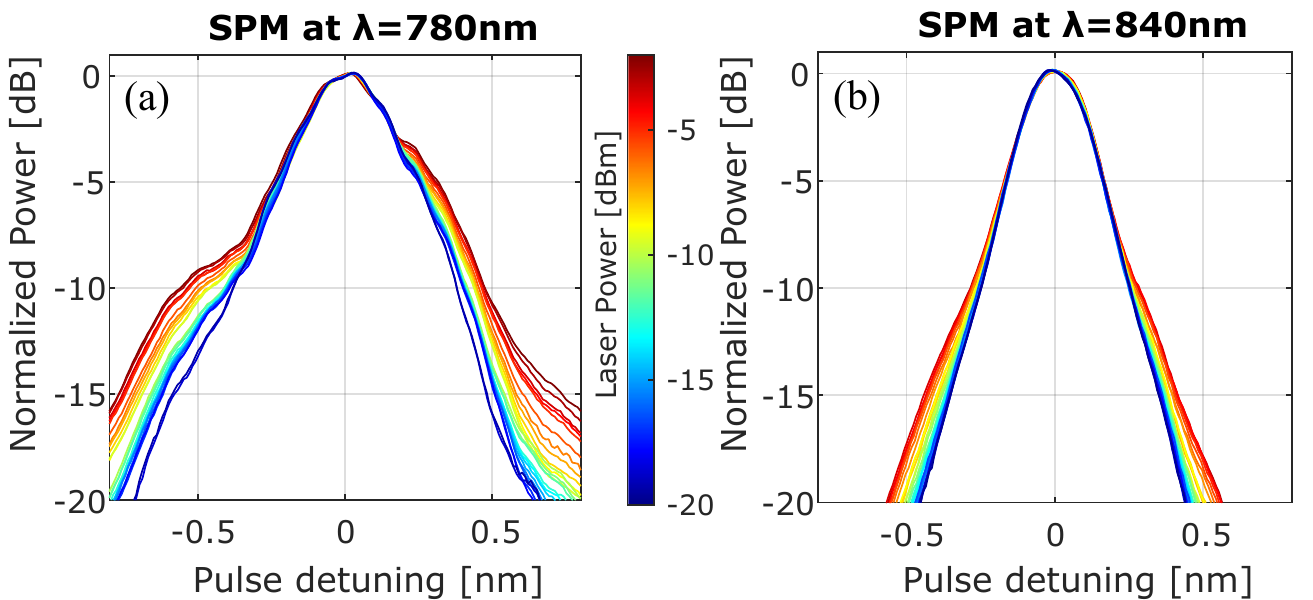}
\caption{Intensity-dependent spectral broadening of an ultra-short laser pulse induced by SPM effect in \ce{SiON} waveguide for two different pulse wavelengths centered at (a) $780$~nm and (b) $840$~nm . The spectra are normalized to their peak powers. The difference in the input pulses lineshapes (the darkest blue lines) at the two central wavelengths is attributed to laser and table optics prior to interaction in the waveguides.}
\label{Fig:F10_SPM}
\end{figure*}

The SPM measurements were obtained using a mode-locked Ti:Sapphire laser, tunable in the wavelength range 720~nm-840~nm, with a $3$-dB pulse-width of $0.2$~nm (2~ps) and a repetition rate of 82~MHz. The laser is directly injected into the \ce{SiON} waveguide using a lensed glass optical fiber. The transmitted pulse is collected with a second identical fiber at the waveguide output and analyzed in an Optical Spectrum Analyzer with $0.04$~nm spectral resolution and sensitivity of~-60~dBm. 

The pulse-broadening experiment was performed for different input powers, ranging from 0~dBm to -20~dBm measured at the output of the waveguide, in order to verify the intensity dependence of SPM.
Considering that the non-linear effects on the input pulse may occur also in the injection optics, composed of lenses and fibers, it is important to attenuate the power injected into the waveguides after the external optics. Therefore, the variation of the power coupled to the waveguide was realized by moving the input-fiber away from the waveguide’s facet, to decrease the coupling efficiency. Figure~\ref{Fig:F10_SPM} shows an example of the experimental data, obtained for two different pulses with central wavelengths at $780$~nm and $840$~nm.

 \begin{figure*}[htb!]
\centering\includegraphics[width=0.66\textwidth]{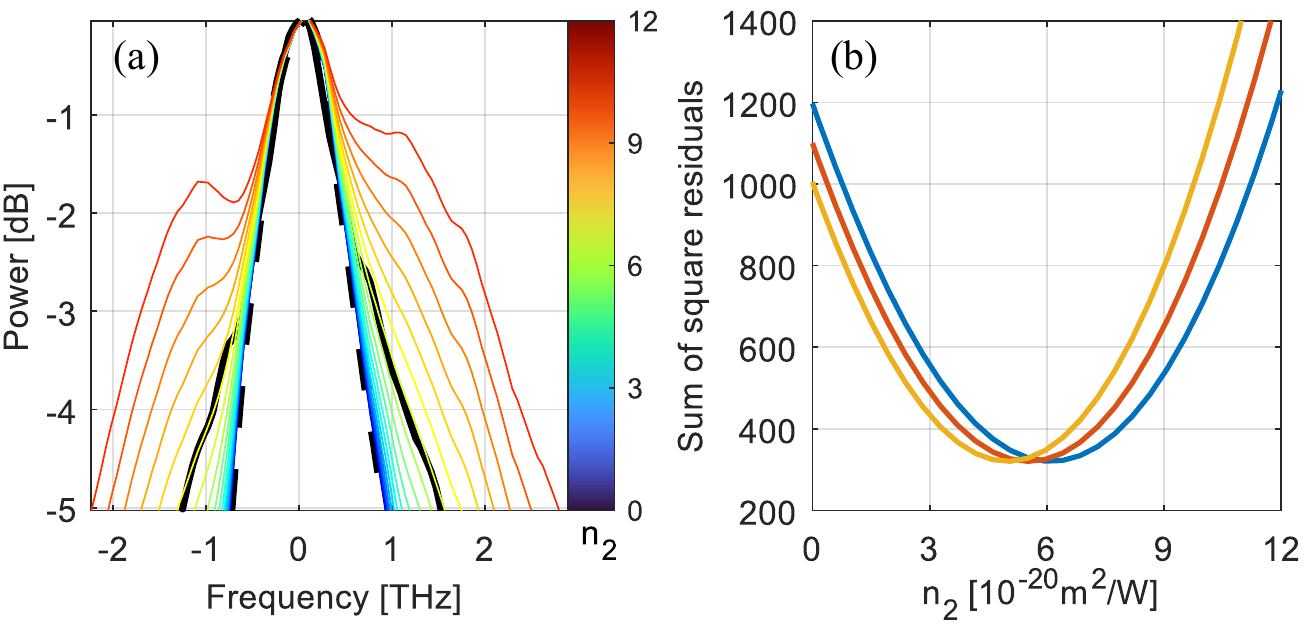}
\caption{Representation of the numerical fitting procedure to estimate the value of $n_2$ at wavelength of $840$~nm. (a)  Starting from the input pulse lineshape (dashed line), a set of output pulse lineshape was simulated for increasing values of the $n_2$ (blue to red, colorbar in units of $10^{-20}$~m$^2$/W), in order to match the experimental lineshape (black solid line). (b) Sum of the square residuals between the theoretical and the measured lineshapes. The minimum position corresponds to the optimal value of $n_2$ that gives the best fit to the experimental data.
Estimations are obtained for three different values of propagation loss $\alpha$, $\alpha+\sigma_\alpha$ and $\alpha-\sigma_\alpha$ (red, blue and yellow, respectively), in order to project the error on the measured loss to the error of estimated $n_2$.}
\label{Fig:F56_n2fiterr}
\end{figure*}

In order to estimate the unknown Kerr nonlinearity $n_2$ of the \ce{SiON} material, we have performed numerical simulations which transform an input pulse spectrum into a broadened one. Having the knowledge of the waveguide’s length and propagation loss, of the effective mode area (obtained by numerical simulations) and of the effective refractive index, the nonlinear Kerr coefficient can be then estimated according to the following procedure. 
For each set of measurements, the lowest-power signal is taken as the reference input pulse-shape. Then, for each power, a set of expected output signals are simulated with the split-step method for different values of $n_2$ (Fig.~\ref{Fig:F56_n2fiterr}a). The final value of $n_2$ is thus estimated as the one that minimizes the spectral difference between the experimental data and the numerical solution (Fig.~\ref{Fig:F56_n2fiterr}b). The main source of error in the estimation of $n_2$ within this approach is given by the error $\sigma_\alpha$ of the propagation loss imposed in the numerical model. The uncertainty of the estimated $n_2$ was evaluated by simulating the spectra at three values of propagation loss: $\alpha$, $\alpha+\sigma_\alpha$ and $\alpha-\sigma_\alpha$, and taking the difference $\delta n_2=(n_2^\textrm{max}-n_2^\textrm{min})/2$ as the estimated error of $n_2$.

\begin{table*}[ht]
    \centering
    \begin{tabular}{c | c c | c c}
    waveguide   & width         & length    & $n_2$ (780~nm)   & $n_2$ (840~nm)     \\ \hline\hline
    wg A        & $1.1 \mu$m    & $27$mm    & $13.1\pm0.5$    & $5.7\pm0.4$       \\  
    wg B        & $1.1 \mu$m    & $37$mm    & $13.5\pm0.5$    & $5.0\pm0.4$       \\
    wg C        & $1.3 \mu$m    & $27$mm    & $12.3\pm0.7$    & $5.8\pm0.5$       \\
    wg D        & $1.3 \mu$m    & $37$mm    & $13.5\pm0.7$    & $5.6\pm0.5$                  
    \end{tabular}
    \caption{Geometrical dimensions and estimated $n_2$ coefficients (in units of $10^{-20}$~m$^2$/W), at two different pump wavelengths, for the different investigated devices. }
    \label{Tab:n2val}
\end{table*}

\begin{figure*}[htb!]
\centering\includegraphics[width=0.66\textwidth]{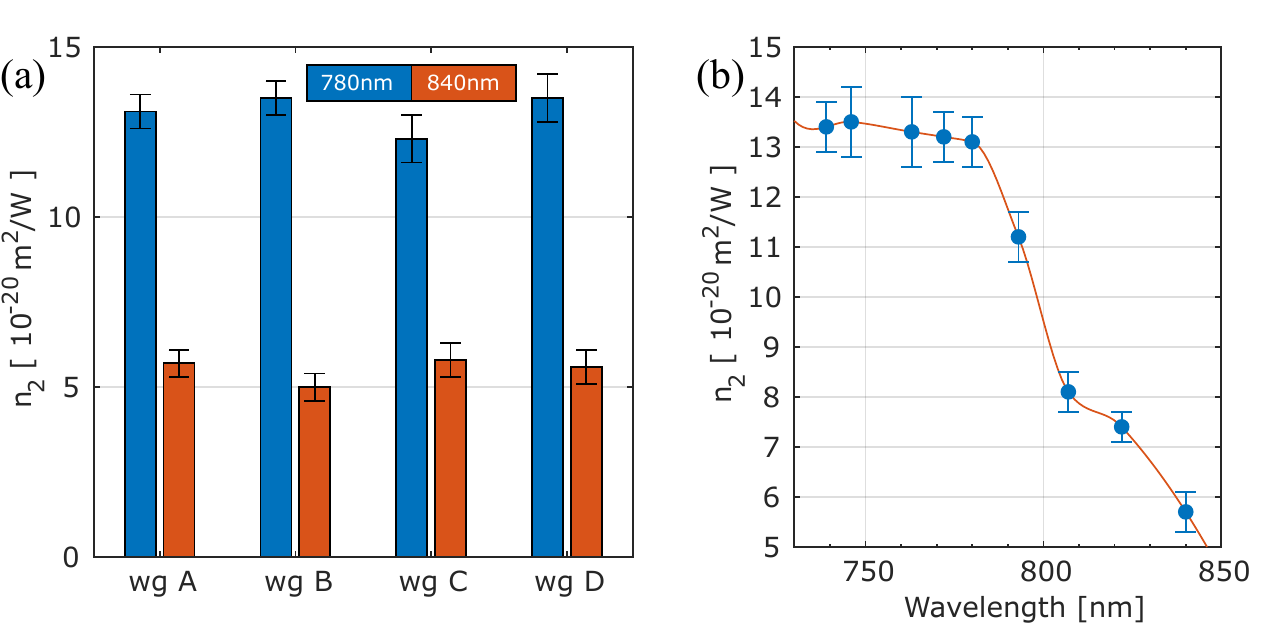}
\caption{(a) Results of $n_2$ estimation for different geometries of waveguides (see Table~\ref{Tab:n2val}) at two different laser wavelengths: $780$~nm (blue) and $840$~nm (orange). The implemented analysis method allows to estimate $n_2$ values which are independent on the waveguide geometry, ensuring that the geometrical properties of the waveguides are weighted properly in the data analysis. (b) The measured spectral dispersion of the nonlinear refractive index $n_2$. An increasing trend towards shorter wavelengths is clearly visible.}
\label{Fig:F78_n2resuls}
\end{figure*}

The described numerical SPM simulations method is based on the experimental inputs, which account for the geometrical dimensions and linear optical properties of the fabricated waveguides. In order to verify the robustness of our approach, we applied this method to study a set of four waveguides. In particular, a pair of waveguides was studied for two different nominal widths of 1100~nm and 1300~nm, as described in Table~\ref{Tab:n2val}.
The validation test, reported in Fig.~\ref{Fig:F78_n2resuls}a, shows that at a fixed wavelength the estimations of $n_2$ are independent on the waveguide's geometry, within error, and therefore confirms the reliability of our analysis. The results are also consistent with previous works~\cite{trenti2018thermopt,ikeda2008thermal,sheik1990dispersion}, indicating that the $n_2$ values of our \ce{SiON} material are in between the expected values for pure \ce{SiO2} and pure \ce{Si3N4}.

In the following, we have selected one of the 1100~nm wide waveguides (sample A) and performed a spectral analysis of the variation of $n_2$ with wavelength. Figure~\ref{Fig:F78_n2resuls}b shows that the Kerr coefficient strongly increases while reducing the pump wavelength from Near-Infrared to Visible-Red region. This behavior is in accordance with the theoretical model that foresees a maximum in the nonlinear coefficient located close to the TPA edge at ~$E_g/2$~\cite{sheik1990dispersion}, corresponding to a wavelength of about $\lambda_\textrm{TPA} \approx 320$~nm for our \ce{SiON} platform.

\section{Dispersion-engineered \ce{SiON} ring resonators for generation of correlated photon pairs}
\label{sec:indexeng}
The knowledge of the linear and non-linear properties of the developed \ce{SiON} photonic platform makes it possible to engineer the modal refractive index, $n_\textrm{eff}(\omega)$, of the waveguide to match particular applications. Nonlinear schemes of generation of non-classical states of NIR photons often relay on SiN  integrated microphotonic devices~\cite{zhao2015visible,cernansky2018complementary,lu2019chip,zhao2020visible}. In this section, we describe our approach for the engineering and realization of \ce{SiON}-based ring resonator devices for on-chip generation of entangled photon pairs via Spontaneous Four Wave Mixing (SFWM).

The dependence of $n_\textrm{eff}$ on the light angular frequency leads to the dispersive nature of the mode's propagation constant $\beta(\omega)= n_\textrm{eff}(\omega)\cdot \omega/c$, which can be expressed as a Taylor series around a central frequency $\omega_0$
\begin{equation}
\beta(\omega)= \beta_0+\sum\limits_{m=1} \frac{\beta_m}{m!} (\omega-\omega_0)^m,
\label{eq:betaTaylor}
\end{equation}
where the diverse orders account for different propagation phenomena. In particular, for nonlinear wave interactions, the first two orders $\beta_1$ and $\beta_2$ play an important role and represent, respectively, the group index $n_g$ and the Group Velocity dispersion (GVD):
\begin{subequations}
\begin{align}
	\beta_1(\omega)=n_g/c=\frac{1}{c} \left(n_\textrm{eff}+\omega \frac{d n_\textrm{eff}}{d \omega} \right),\\
	\beta_2(\omega)=\frac{d\beta_1}{d\omega}=\frac{1}{c} \left(2\frac{d n_\textrm{eff}}{d \omega} + \omega \frac{d^2 n_\textrm{eff}}{d \omega^2} \right).
\end{align}
\label{eq:GVD}
\end{subequations}

In a ring resonator, the group velocity $\beta_1$ defines the spectral separation $\delta_c$ between successive resonant modes -- the \textit{free-spectral range} (FSR) -- following the relation $\delta_c=2\pi/(\beta_1 L_c)$, where $L_c$ is the physical length of the cavity. A flat dispersion of $\beta_1(\omega)$ over a range of frequencies provides energy-equidistant cavity resonances, which is essential to fulfill the energy conservation requirement, for example, in nonlinear FWM and frequency comb generation experiments~\cite{cernansky2018complementary,zhao2020visible}, where the pump, the signal and the idler should satisfy the relation $\omega_i = 2\omega_p - \omega_s$.

\begin{figure*}[htb!]
\centering \includegraphics[width=0.66\textwidth]{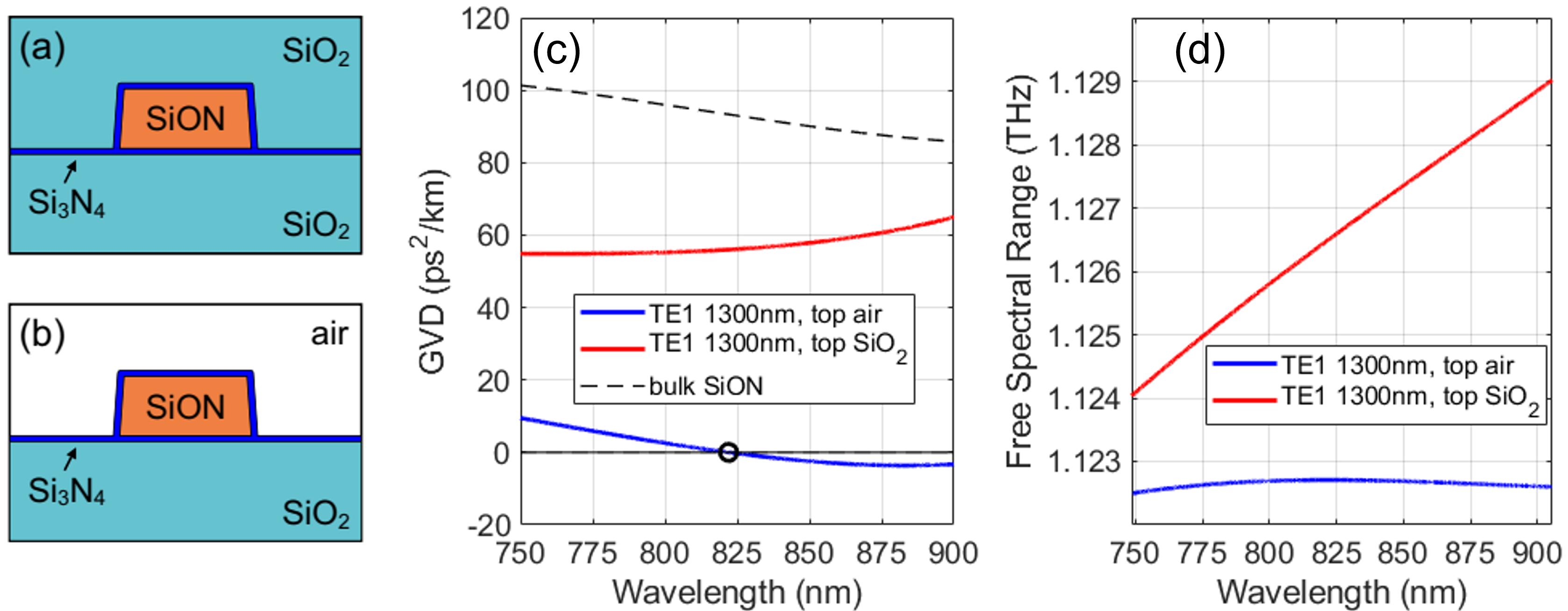}
  \caption{(a) Sketch of the cross-sectional geometry of the \ce{SiON} waveguide embedded in SiO$_2$ matrix. (b) The same geometry with the top SiO$_2$ cladding substituted with air. (c) The calculated GVD's of quasi-TE-modes for the respective geometries, showing that anomalous GVD is achieved when the top SiO$_2$ cladding is substituted with air.  For comparison, the dispersion of the bulk \ce{SiON} material is also shown (dashed line). (d) The calculated FSR of two rings evidence the sufficiently flat $\delta_c$-trend for the air-cladded device.}
  \label{fig:geom}
\end{figure*}

The parameter $\beta_2$ describes how the different spectral components of a propagating pulse travel, and in either case $\beta_2>0$ (\textit{normal} dispersion) or $\beta_2<0$ (\textit{anomalous} dispersion) result in temporal broadening of the pulse. The case of $\beta_2=0$ at some frequency -- the zero dispersion frequency (ZDF) -- is of particular interest for nonlinear optical applications because around the ZDF different spectral components experience largely reduced second-order dispersion. In SFWM experiments with microring resonators around ZDF the spectral spread of $\delta_c$ is minimal and the flatness of $n_g$ provides with larger nonlinear generation bandwidth~\cite{agha2009theoretical,agrawal2000nonlinear}.

We performed numerical axisymmetric simulations based on the Finite Elements Method in order to develop proper geometries for a device operating at NIR wavelengths, supporting single-mode characteristics and showing an anomalous GVD over the spectral range of interest. The radius of the studied ring resonators was set to 25~$\mu$m. This choice is motivated by the necessity to keep the ring radius large enough to avoid radiative losses but small enough to provide with $\delta_c\sim2.5$~nm in order to minimize the spectral overlap of pump pulses with more than one mode around 800~nm of wavelength.

We first investigated a conventional geometry, in which the waveguide of a slightly trapezoidal form is fully embedded within the SiO$_2$ cladding (Fig.~\ref{fig:geom}a). The lateral boundaries of the waveguide have an inclination angle of $86^o$, which is the typical value according to our fabrication process. The height of the \ce{SiON} waveguide was fixed to 500~nm. Simulations were performed by varying the ring waveguide width from 900~nm to 1700~nm and the azimuthal number $M$ of the fundamental radial mode from 270 to 330, covering a wavelength span of 200~nm around $\lambda=850$~nm. The obtained results for both the quasi transverse electric (TE) and magnetic (TM) modes showed that the GVD remains normal for all ring widths within the spectral range between 750~nm and 950~nm.

\begin{figure*}[htb!]
\centering \includegraphics[width=0.66\textwidth]{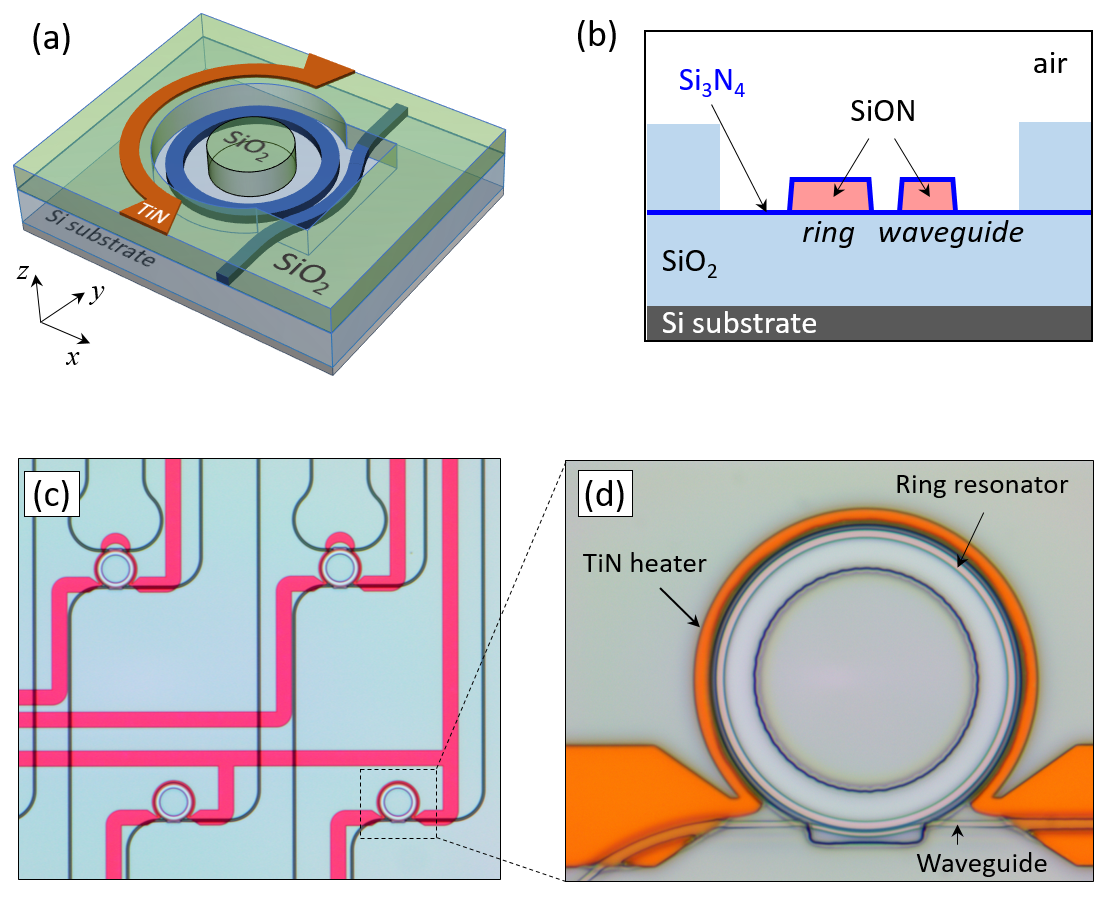}
  \caption{(a) A 3D sketch of the ring resonator with top-air cladding and a TiN phase shifter. (b) Cross-sectional geometry at the ring-waveguide coupler position. (c) Optical image of the fabricated chip with arrays of ring resonators and (d) the zoom around a ring resonator in all-pass configuration.}
  \label{fig:ring}
\end{figure*}

Next, we performed simulations of the same ring geometry substituting the top SiO$_2$ cladding with air (Fig.~\ref{fig:geom}b). This new configuration turned to be particularly interesting since the GVD appeared to be much more sensitive to the variations of the waveguide width. We found that the GVDs of air-cladded resonators are already anomalous for smaller widths ($\sim 1100$~nm) and turn slowly into a normal one over 1350~nm. In Fig.~\ref{fig:geom}c we compare the calculated GVDs for 1300~nm waveguide width rings in both configurations with top SiO$_2$ (red) and air-claddings (blue). A zero-dispersion point is observed at around $\lambda\approx 822$~nm for a 1300~nm waveguide width rings. The air-cladded ring shows sufficiently flat dispersion of the group index and, consequently, a stable FSR over a large span of wavelengths from 800 to 850~nm's (Fig.~\ref{fig:geom}d). The effective nonlinearity $\gamma$ for the $1.3\times0.5~\mu$m$^2$ ring at the zero-dispersion point of 822~nm is estimated to be of the order of 1~m$^{-1}$W$^{-1}$.

A schematic of the dispersion-engineered resonator, which has been fabricated on the same chip together with spiral waveguides, is shown in Fig.~\ref{fig:ring}a. The top SiO$_2$ cladding around the resonators was removed via a selective wet-etching procedure in buffered HF solution during which the 50~nm Si$_3$N$_4$ film acts as an ideal etch-stop film to prevent underetching of waveguiding components. The opening window around the ring was designed such that the remaining top SiO$_2$ cladding after wet etching is retracted to a safe distance of at least $2.5~\mu$m from the waveguide (Fig.~\ref{fig:ring}b). Figure~\ref{fig:ring}c shows an optical micrograph of arrays of ring resonators both in add-drop and all-pass configurations, while a blowup around an air-cladded all-pass ring resonator is shown in Fig.~\ref{fig:ring}d. The linear characterization and non-linear sFWM experiments on these devices are currently ongoing, however, their discussion is out of the scope of the current manuscript.


\section{Conclusions}
\label{sec:concl}
In this work, we have demonstrated a new silicon oxynitride-based integrated photonic platform for linear and nonlinear application in the VIS-NIR wavelength range.
The fabricated devices show a low propagation loss $<2$~dB/cm, comparable to commercially available devices at $800$~nm wavelength and with ongoing improvement of at least a factor two.
We have demonstrated that, despite the reduction of the refractive index with respect to \ce{SiN}, the \ce{SiON} waveguides preserve a relatively strong optical nonlinearity of $13\pm06 \times 10^{-20}$~m$^2$/W around the wavelength $780$~nm. Furthermore, thanks to the possibility to remove locally the cladding without damaging the waveguide, our platform allows for a larger versatility in engineering the waveguide dispersion. This enables to investigate one of the specific and peculiar applications of nonlinear FWM by properly adjusting the group index and group velocity dispersion in order to enhance nonlinear photon pair generation in ring resonators.
By combining the good linear properties, the promising optical nonlinearities for on-chip photon generation and our recently developed technology for on-chip photon detection~\cite{bernard2021top}, we envision the potential of this platform to achieve, in the near future, a full integration of photon generation sources, manipulation and detection on a single Silicon chip, operating at room temperature, for classical and quantum photonics applications.


\section{Acknowledgment}

\noindent\textbf{Funding.} European Commission (777222, 899368).

\noindent\textbf{Acknowledgments.}
The authors are thankful to M.~Bernard and G.~Pucker for the fruitful discussions regarding the design and fabrication of the photonic devices.\\
\noindent\textbf{Disclosures.}
The authors declare no conflicts of interest.

\noindent\textbf{Data availability.} Data underlying the results presented in this paper are not publicly available at this time but may be obtained from the authors upon reasonable request.


\bibliography{SiON_PIP_ref}

\end{document}

%% file: preamble.tex
\usepackage{amsthm}
\usepackage{mathtools}
\usepackage{physics}
\usepackage{xcolor}
\usepackage{graphicx}
\usepackage[left=18mm,right=18mm,top=35mm,columnsep=15pt]{geometry} 
\usepackage{adjustbox}
\usepackage{placeins}
\usepackage[T1]{fontenc}
\usepackage{lipsum}
\usepackage{csquotes}
\usepackage[version=3]{mhchem}